%% file: main.tex
\documentclass[sigconf]{acmart}
\acmConference[SE 2030]{International Workshop on Software Engineering in 2030}{November 2024}{Puerto Galinàs (Brazil)}

\AtBeginDocument{%
  \providecommand\BibTeX{{%
    \normalfont B\kern-0.5em{\scshape i\kern-0.25em b}\kern-0.8em\TeX}}}

\setcopyright{acmlicensed}
\copyrightyear{2018}
\acmYear{2018}
\acmDOI{XXXXXXX.XXXXXXX}

\usepackage{import}
\usepackage{multirow}
\usepackage{graphicx}

\usepackage{dirtytalk}

\begin{document}

\title{Exploring the Experiences of Experts: Sustainability in Agile Software Development - Insights from the Finnish Software Industry}

\author{Hatef Shamshiri}
\authornote{All authors have contributed equally to this research.}
\orcid{0000-0002-6048-7995}
\affiliation{%
  \institution{LUT University}
  \city{Lappeenranta}
  \country{Finland}
}
\email{hatef.shamshiri@lut.fi}
\author{Ashok Tripathi}
\affiliation{%
  \institution{LUT University}
  \city{Lappeenranta}
  \country{Finland}
}
\email{ashok.tripathi@lut.fi}
\author{Shola Oyedeji}
\affiliation{%
  \institution{LUT University}
  \city{Lappeenranta}
  \country{Finland}
}
\email{shola.oyedeji@lut.fi}
\author{Jari Porras}
\affiliation{%
  \institution{LUT University}
  \city{Lappeenranta}
  \country{Finland}
}
\email{jari.porras@lut.fi}
\renewcommand{\shortauthors}{Shamshiri et al.}

\begin{abstract}
\import{sections}{abstract}
\end{abstract}



\received{20 February 2007}
\received[revised]{12 March 2009}
\received[accepted]{5 June 2009}

\maketitle

\import{sections}{introduction}

\import{sections}{background}

\import{sections}{methods}
\import{sections}{results}
\import{sections}{discussion}
\import{sections}{validity}
\import{sections}{roadmap}
\import{sections}{conclusion}

\section{Acknowledgments}

The Academy of Finland funded this study under the ICT for Climate Actions project (23B3C034YT10).

\bibliographystyle{ACM-Reference-Format}
\bibliography{zotero}
\import{sections}{appendix}
\end{document}

%% file: sections/abstract.tex
Agile software development is gaining popularity among software developers due to its benefits. As the interest in agile software development grows, there is an increasing focus on investigating sustainability within this field. This study aimed to explore sustainability within agile software development in the Finnish software industry and, through gathered experiences, contribute to the software engineering roadmap 2030. Using an interview approach, we conducted an empirical study within the Finnish software industry to achieve this goal. The findings indicate a growing interest among experts in integrating sustainability into agile software development. The results show that the Scrum methodology is the most popular approach in the Finnish software industry, and addressing different sustainability dimensions can have a ripple effect on each other. The study proposes three key elements to be considered in the software engineering roadmap 2030: integrating sustainability into software engineering education, creating sustainability tools and frameworks, and assessing the energy efficiency of libraries used in software development. 

%% file: sections/introduction.tex
\section{Introduction}
The UN member states incorporated the 2030 agenda for sustainable development, highlighting sustainable development goals that aim to promote sustainable development between 2016 and 2030 \cite{sdg_agenda_nodate}. Software products and/or services are an omnipotent facet of today's life, and consequently it becomes imperative that software development and usage be aligned to the sustainable agenda as well.
At present a greater common understanding of software sustainability is lacking, with the software engineering community, individuals, groups, and organizations holding divergent viewpoints. Amsel et al. \cite{amsel_toward_2011} suggest sustainable software engineering aims to develop reliable, durable software that meets users’ needs while reducing its environmental impacts. Organizations need to adopt practices that promote sustainability within their software development processes.

A popular process/methodology that software organizations are increasingly adopting in software development is Agile. According to the Agile Manifesto, agile processes enable sustainable development, and the stakeholders, including sponsors, developers, and users, must maintain a sustainable pace \cite{beck_manifesto_2001}. The agile manifesto emphasizes frequent software delivery, embracing changes even late in development, and face-to-face communication. According to a study, agile methods positively impact the software development life cycle, increasing developers’ productivity and leading to more energy-efficient software \cite{rashid_developing_2016}. Agile can play a crucial role in sustainability by enabling teams to deliver more efficient, reliable, and environmentally friendly software \cite{dick2013green} \cite{pankowska2013sustainable} \cite{tate2005sustainable}. 

The current study investigates challenges, impacts, and approaches to agile software development in the context of sustainability in the software industry. This includes understanding the challenges in agile software development that affect sustainability, how these challenges impact agile software development, and what could be improved to achieve more sustainable agile software development. 
The objective of this study is to develop a roadmap for more sustainable agile software development within the software industry by conducting an interview study.

%% file: sections/background.tex
\section{Background}
The definition of sustainability in software differs considerably from one publication to the other, so it is necessary to identify a focal point on which sustainability in software may be addressed. An example definition is introduced by Sedano et al. \cite{sedano_sustainable_2016}, defining sustainable software development as the capacity and willingness of a software development team to minimize the adverse effects of substantial disruptions, especially team turnover, on productivity and effectiveness. Penzenstadler et al. \cite{penzenstadler_generic_2013} defined the five dimensions of sustainability as outlined below.
\begin{itemize}
    \item \textbf{Technical}: Technical sustainability aims to ensure the long-term use of systems and an adequate evolution in response to changing surroundings and requirements.
    \item \textbf{Social}: Social sustainability means maintaining social capital and preserving solidarity within social communities.
    \item \textbf{Environmental}: Environmental sustainability aims to protect natural resources to improve human welfare.
    \item \textbf{Economical}: The objective of economic sustainability is to maintain assets.
    \item \textbf{Individual}: Sustaining the value of an individual’s human capital involves maintaining the personal good of the individual.
\end{itemize}
The pervasive role of software in today's era means that questions about its development and usage cannot be ignored. Throughout the software life cycle, green and sustainable software should have a minimal adversarial impact on the economy, society, and the environment. The concept of green and sustainable software engineering entails implementing sustainable software by following a green and sustainable software engineering process \cite{dick2013green}. 

Lago \cite{lago2015challenges} suggests that numerous publications have been published in the field of green software; most of them, however, offer only partial or no information on topics, e.g., developing energy-efficient software and determining the relevant metrics and tools. Achieving sustainability awareness in software system engineering requires incorporating the aspects of sustainability into the entire software engineering life cycle and the underlying business processes \cite{betz_sustainable_2014}.

%% file: sections/methods.tex
\section{Research Approach}
This research aimed to investigate the software industry's status on sustainability in agile software development and 
contribute to software engineering roadmap. 
We have conducted a structured interview study among the Finnish software industry experts to understand sustainability challenges, impacts, and ways to improve sustainability in agile software development. We have designed a fixed set of questions to ask the participants based on the current study's research questions.
The researcher acted as an interviewer while letting the participants answer the questions freely without interruption. To avoid researcher bias during the interviews, except for clarifying questions, the researcher only listened and avoided expressing his narrative. We have detailed the research approach into different subsections below.

\subsection{Research Questions}

In this study, we are investigating sustainability in agile software development in the Finnish software industry. We look at and answer the following research questions (RQ) in this interview study.

 \begin{itemize}
     \item \textbf{RQ1:} What is the perception of sustainability in the software industry utilizing agile practice(s)?
     \item \textbf{RQ2:} What are the challenges affecting sustainability in agile software development?
     \item \textbf{RQ3:} How do the sustainability challenges impact agile software development?
     \item \textbf{RQ4:} What could improve sustainability in agile software development?
 \end{itemize}

\subsection{Sampling}

Fifteen experts from seven different organizations participated in this study. 
The experts had 2 to 18 years of experience in their current roles in their organization and 3 to 40 years of experience in software development in general. They had many different roles, including but not limited to corporate responsibility manager (CRM), chief growth officer (CGO), head of research, agile coach, staff expert of sustainability, team leader, product development lead, scrum master, green code expert, and product owner (PO). The participating organizations operate in industrial areas such as telecommunication, ICT, IT, software services, and consultation in Finland. Most participating organizations are categorized as large, with more than 250 employees, while few are categorized as mid-sized organizations with less than 250 employees. The organizations were found among the author networks and through web search using the criteria of software businesses operating in Finland. The expert's roles are outlined in Table \ref{tab:practitioners}.

\begin{table}[H]
\caption{Experts roles.}
\label{tab:practitioners}

\begin{tabular}{|l|l|}
\hline
\textbf{ID} & \textbf{Job Title}            \\ \hline
E1&Team manager              \\ \hline
E2 & Business design                           \\ \hline
E3  &CRM                                     \\ \hline
E4  &Manager                   \\ \hline
E5  &Technical manager, Scrum master, Software developer                                  \\ \hline
E6  &CGO, green code expert   \\ \hline
E7  &PO, Scrum master   \\ \hline
E8  &Team leader, architect  \\ \hline
E9  &Agile coach \\ \hline
E10  &Head of research  \\ \hline
E11  &Agile coach  \\ \hline
E12  &Senior designer, sustainability staff expert  \\ \hline
E13  &Portfolio lead  \\ \hline
E14  &Product development lead  \\ \hline
E15  &Devops and cloud engineer \\ \hline
\end{tabular}%

\end{table}
\subsection{Interviews}

We conducted interviews using mostly open-ended questions but also a few closed-ended questions, allowing experts to express their opinions freely and justify their answers even when it comes to closed-ended questions (see Appendix). The questions were defined to enable experts to express their views freely about the topic within their organization. The questions were designed to fulfill this study's objective related to sustainability in agile software development in software companies operating in Finland. 

\begin{itemize}
    \item \textbf{Demographics}. For the initial interview questions, we were interested in the experts' roles and experiences and the size and industry of their respective organizations. We asked about their job roles, years of experience, size, and industry in their respective organizations.
    \item \textbf{Methods}. We asked the experts about agile methods utilized in their respective organizations.
    \item \textbf{Sustainability perceptions}. For the second part of the interview, we were interested in the experts and their respective organizations' perceptions of sustainability in software development in general and, more specifically, in agile software development. We asked them about their perceptions of sustainability in software development, sustainability dimensions, sustainability considerations in their organization, and organizational sustainability goals relevant to agile software development.
    \item \textbf{Agile sustainability}. We asked the experts about challenges in agile software development that affect sustainability, how those challenges impact agile software development, and what they think can be done to improve sustainability in agile software development.
\end{itemize}
The interviews in the study were conducted either in person or online, such as through Zoom and Google Meetings. The meetings were recorded with the experts' explicit consent, and they were presented with the terms and conditions of the recording, including who would transcribe the interview and the anonymity of the experts and participating organizations.

\subsection{Data Analysis}

We have analyzed the study results using qualitative approaches to answer the research study questions. We have utilized NVivo (software intended for qualitative data analysis) to transcribe and code the data and Excel to visualize and analyze the information. We have used a thematic analysis approach to establish themes and patterns to analyze the data further. The thematic analysis approach helps us develop patterns and themes that aid us in understanding the collected data and assessing them. We have categorized data into four categories: demographics, methods, perceptions, and agile sustainability. We have prepared the data collected from the interviews for further analysis, transcribing the interviews to extract the relevant data. We have further anonymized data to adhere to the confidentiality terms and conditions of the interview study and to safeguard the identity of the participants and organizations. We have used an inductive approach to assess the obtained data and developed themes based on the nature of the data to form themes for further study. 

%% file: sections/results.tex
\section{Study Results}

This section details the study's results. It is structured as follows: Agile Methods explains current methods used in organizations; Perception details results related to software sustainability; Dimensions explains considerations for organization sustainability; and finally, Agile Sustainability details challenges and approaches related to sustainability in agile software development.

\subsection{Agile Methods}

\begin{figure}[h]
  \centering
  \includegraphics[width=1\linewidth]{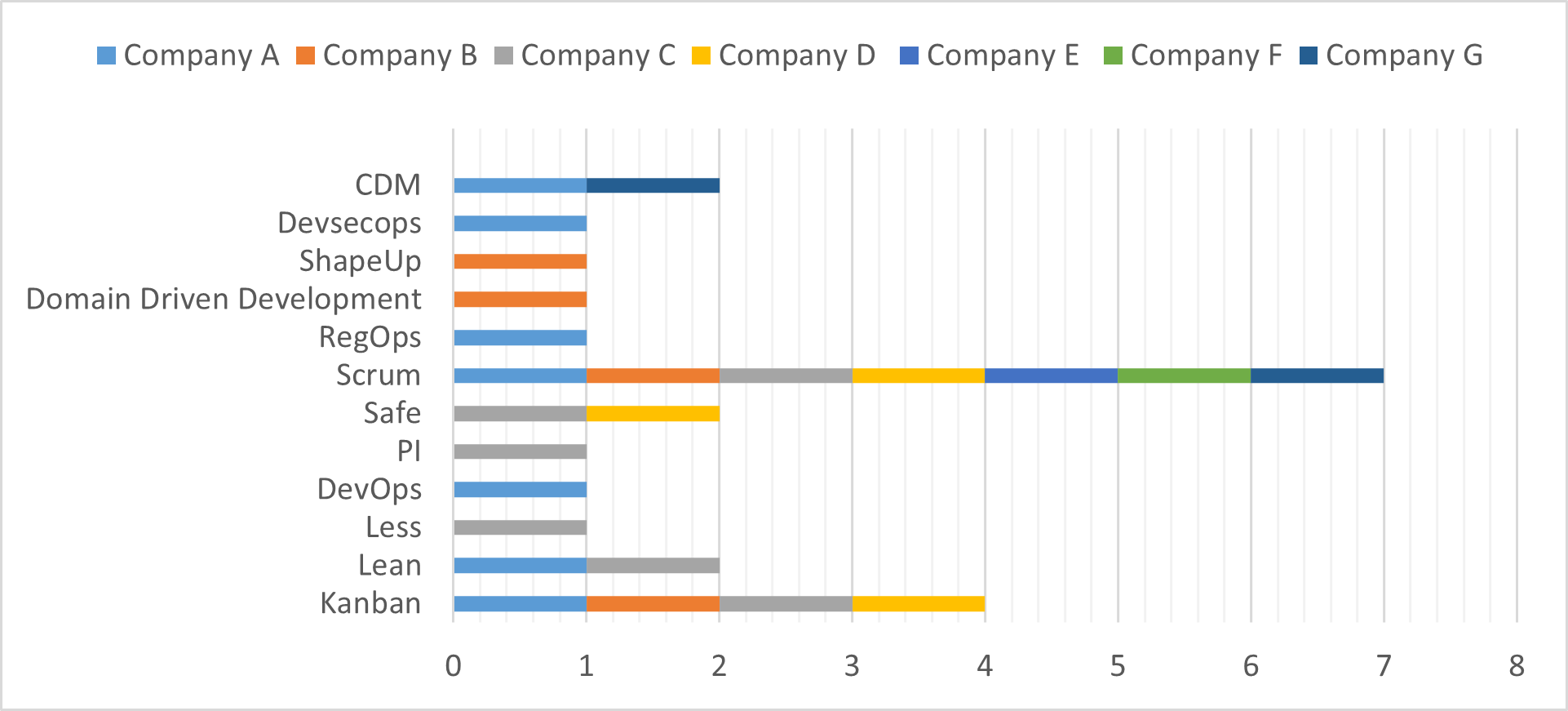}
  \caption{Agile methods utilized by the participating organizations.}
  \label{agilemethods}  
\end{figure}

Most experts indicated that their organizations follow different agile methods, as shown in Figure \ref{agilemethods}. Among the agile methods presented, Scrum has been among the most utilized by organizations. Some organizations have either developed their version of an agile approach or customized the agile methods to meet their specific needs, which we call the Company Devised Method (CDM) in this paper. However, the names and nature of these methods are withheld to adhere to the confidentiality of the participating organizations.

\subsection{Perception}
In this section, we detail results related to software sustainability, sustainability dimensions, and sustainability considerations in organizations.
\subsubsection{Software Sustainability}

We asked the experts about their perception of sustainability in software development. They suggested that sustainability is a new and underrepresented topic and that most developers lack awareness of certain aspects of sustainability, such as energy consumption in software development. They suggested that most developers lack education regarding software's energy consumption and carbon emissions. Developers agree that something must be done when presented with the issues, but there is uncertainty regarding whether they will act or expect someone else to. The experts suggested that everyone should care about and act on sustainability. Experts show significant concern regarding software energy usage and carbon emissions. "I am concerned about how they use the energy or the electricity, the applications; the less they use, the better." (E8).   

A significant concern among experts was the energy usage of software products and the carbon emissions of digital and cloud services, product delivery, etc. They suggested that reducing the energy usage of software and emissions of product delivery and the whole value chain is essential. They also considered other sustainability dimensions, such as technical, social, individual, and economic, critical. Software performance should continuously improve, green coding practices should be adopted, and software durability and accessibility should also be improved according to them. Experts also highlighted that a sustainable pace should be maintained among developers, as suggested by the Agile manifesto \cite{beck_manifesto_2001}, to avoid burnout.

\subsubsection{Dimensions}
\begin{figure}[h]
  \centering
  \includegraphics[width=1\linewidth]{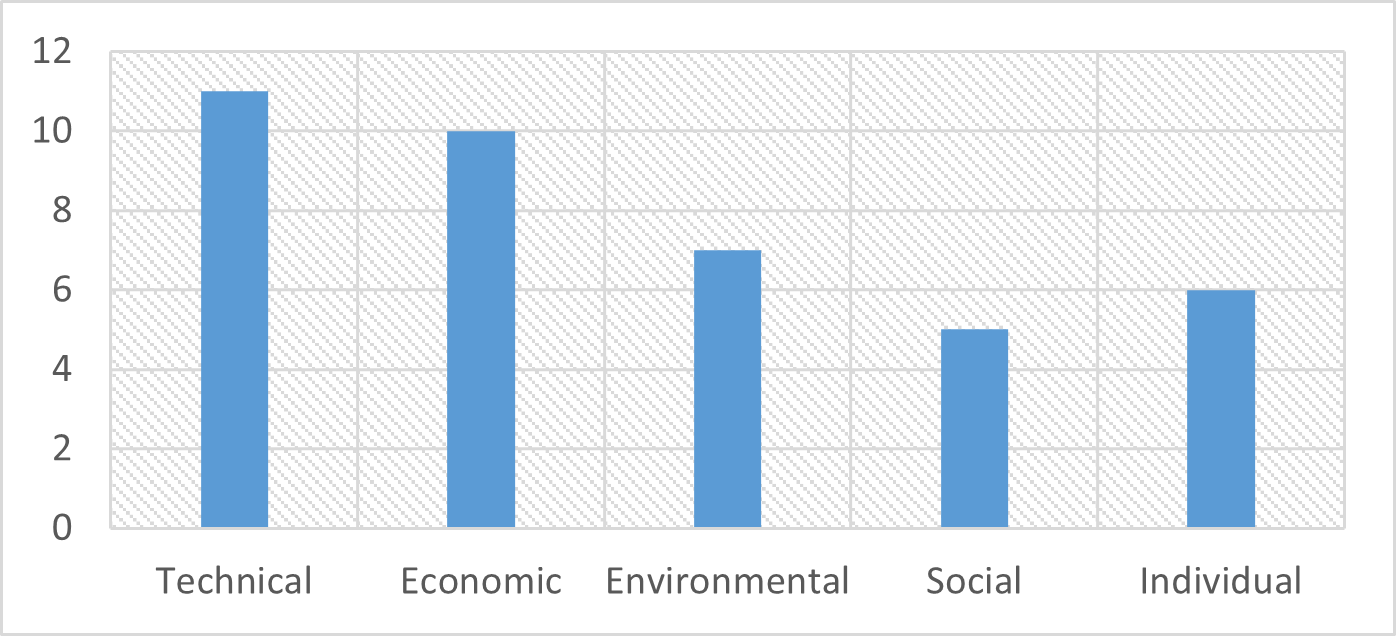}
  \caption{Sustainability dimensions.}
  \label{dimensions}
\end{figure}

We have asked the experts to select the top three sustainability dimensions most relevant to agile software development. They ranked technical, economic, and environmental dimensions among the five sustainability dimensions as the top three, as observed in Figure \ref{dimensions}. They also suggested that impact on one sustainability dimension can impact another. "When you make software which has more performance, you are also making software that doesn't need that much infrastructure to run, so it is an economic sign" (E1).  According to the experts, software that accommodates users' needs while omitting unneeded features can positively impact environmental sustainability. Therefore, according to the experts, the technical, economic, and environmental sustainability dimensions are most relevant to agile software development.

Although most experts considered the dimensions above the most relevant ones, they also showed concern for individual and social dimensions. An expert mentioned that social sustainability is significant in Scrum as, according to the expert, it is all about collaboration and giving voice to people who would stay silent otherwise, such as in Scrum's daily practices. According to the expert, Scrum's daily practices provide a safe spot for stakeholders to communicate. Furthermore, experts considered well-being an aspect of individual sustainability important in agile software development, stating that individuals who do well make better products.  
\subsubsection{Sustainability Consideration in Organizations}
We asked the experts to rate the sustainability considerations within the organization on a scale of 1, meaning lowest, to 5, meaning highest. Most experts indicated that their organizations consider sustainability highly yet still require improvement, so no one gave the highest possible score, 5, to the sustainability consideration. Most rated sustainability consideration above 3, while a few gave equal or below that score. Furthermore, the experts were asked their opinions regarding their organizations' sustainability goals. 

The experts provided goals that relate mainly to four sustainability dimensions: technical, individual, social, and environmental. They highlighted goals such as continuous learning and self-development, accessibility, performance analysis, continuous improvement, carbon neutrality, awareness, and a company culture that ensures employees' work does not harm society and the environment. "We want our employees to keep learning new things as new technologies and ways of working models emerge" (E13). "Accessibility is taken into account in design and development, and then they begin testing it, so it's one of the sort of streams inside the agile that everything that we do needs to be accessible" (E6). They considered aspects mentioned of importance to agile software development. However, some experts indicated that their organization has sustainability goals, yet actions still needed to be set to achieve them, and few stated that their organizations lack such goals.

\subsection{Agile Sustainability}
An expert considered misinformation and misunderstandings regarding agile software development affecting sustainability. According to the expert, the challenges mentioned burden communication due to a lack of shared understanding of the terms and iterative way of working in agile. Furthermore, according to another expert, a lack of shared understanding hinders teams from following the same sustainable principles in different projects. Another challenge suggested by the experts includes a lack of information about the energy efficiency of libraries they may use in software development. They have shown concerns regarding software quality and customer value challenges.

One of the challenges suggested was technical debt, as it slows the development process and burdens agile software development. The experts indicated that mitigating technical debt is not visible to customers and needs to produce tangible value for them, which the experts consider to be challenging. 
They also suggested that adding green coding and accessibility is constrained as it increases costs, and customers may not see value in implementing them. Another constraint indicated by the experts is related to time constraints affecting the economic sustainability of the process. Sometimes, problems arise in projects that need to be considered initially, which lengthens the development process. The experts indicated they could not bill the extra hours spent fixing that problem, negatively affecting economic sustainability. The challenges mentioned by the experts are detrimental to agile software development's technical and economic sustainability. 

We asked the experts their opinions on what can be done to improve sustainability in agile software development. They suggested that everyone should take responsibility for the sustainability of their work, so it is vital to have a checklist for sustainability for individuals to help them achieve sustainability. The experts highlighted the importance of empowering individuals and, consequently, empowering teams and establishing a sustainability mindset among individuals. A practice indicated by the experts was to bring up topics such as sustainability in weekly or planning sessions to remind everyone of sustainability and keep sustainability active in their memory, influencing their daily decision-making. 

%% file: sections/discussion.tex
\section{Discussion}
Sustainability is a relatively new topic in software and, according to the experts, an underrepresented one; therefore, we must embody sustainability within software engineering education. For sustainability to become an integral part of the software development processes, the software developers must possess sustainability knowledge. According to the experts, software developers need proper education, which may stem from a lack of inclusion of sustainability in software engineering studies. An approach to promote a sustainability mindset among software developers is to include sustainability within software engineering education. Adopting sustainability within software engineering education can improve awareness and knowledge of sustainability among future software practitioners. There is a need to develop strategies and policies to increase sustainability awareness and knowledge among stakeholders within the software industry. 

\begin{figure}[h]
  \centering
  \includegraphics[width=1\linewidth]{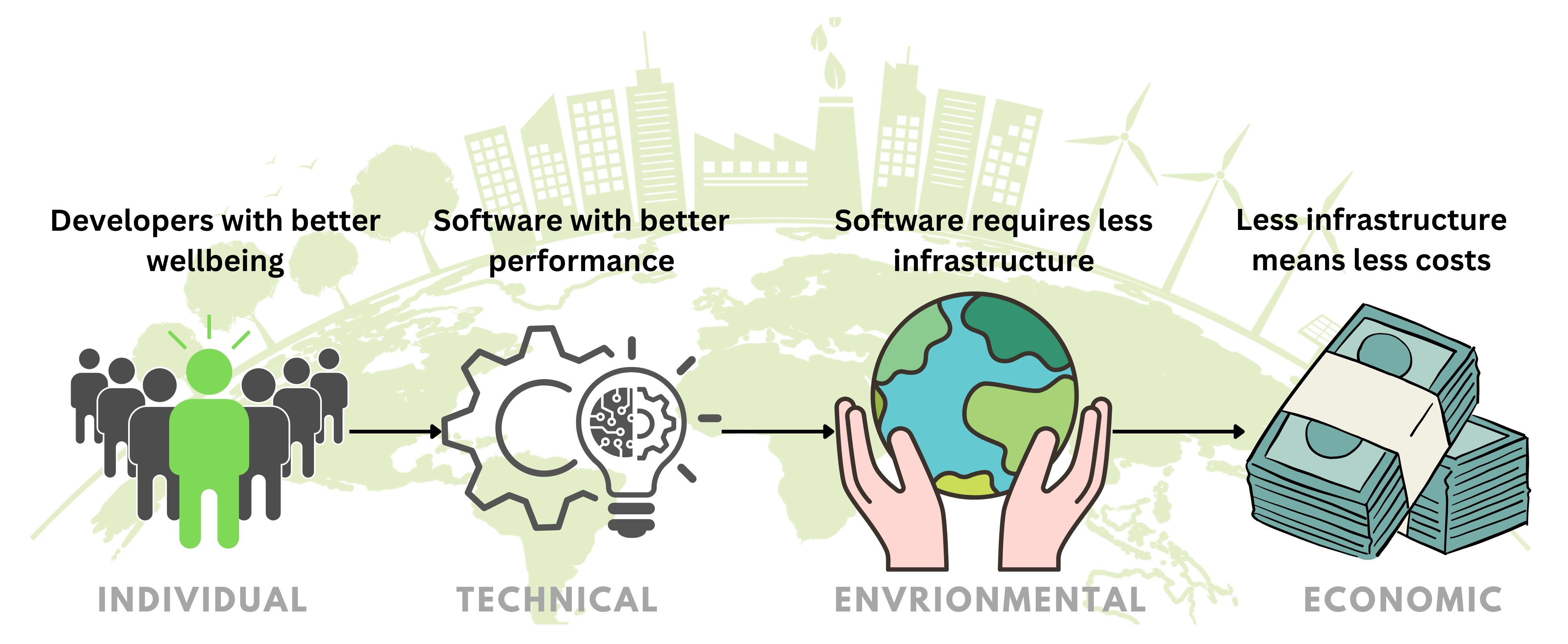}
  \caption{Sustainability dimensions chain effects.}
  \label{dimensionseffects}
\end{figure}

 The development of tools and frameworks is necessary to improve sustainability in agile software development. A sustainable agile software development must address different sustainability dimensions, such as technical, environmental, and economic. From experts' responses, addressing software's technical sustainability can positively impact environmental and economic sustainability. It suggests that sustainability among different dimensions has a chain effect on each other within agile software development. Furthermore, experts considered well-being an essential aspect of individual sustainability, stating that individuals who do well make better products, impacting technical sustainability as shown in Figure \ref{dimensionseffects}. According to the study results, social sustainability is critical to agile software development, particularly Scrum, as it is about collaboration and communication. A significant challenge to social sustainability is misinformation and a lack of shared understanding in agile software development, a considerable communication concern. Developing new approaches and improving existing ones is necessary to enhance communication among agile teams. It is critical to develop tools and frameworks that aid organizations in assessing and strengthening sustainability within their agile software development processes. 

The libraries used in agile software must be measured for their efficiency. More information about libraries used in software development, such as their energy efficiency, needs to be provided, according to the study results. The lack of information about the energy efficiency of libraries used in software development can lead to adverse impacts on the environmental sustainability of software products. For example, practitioners show concern regarding the energy efficiency of large language models (LLM). An LLM is an artificial intelligence (AI) system capable of understanding and producing text, among other functions \cite{LLM_what_CLF}. These models are trained on extensive datasets, so they are termed "large." The popularity of LLMs is on the rise, and understanding different LLMs' energy efficiency is critical for improving agile software development environmental sustainability. There is a need to measure the energy efficiency of libraries used in software development and LLMs to aid practitioners in choosing the most energy-efficient option in agile software development. 

When implementing sustainability approaches, we should prioritize agile methods that can have a broader impact within the software industry. Organizations participating in this study followed multiple agile methods, and Scrum was the most commonly utilized agile method among them. It indicates Scrum’s popularity among Finnish software companies; therefore, addressing sustainability in the mentioned practice can have a more widespread impact. It is a positive step toward a more sustainable future to develop approaches to improve sustainability within agile software development, particularly Scrum, which can considerably impact sustainability. Interest in sustainability is growing in the Finnish software industry, as evidenced in the study results, and addressing it can have a significant positive impact. 

%% file: sections/validity.tex
\section{Study Validity}
We adhere to the following criteria in our study \cite{easterbrook_selecting_2008}.

\textit{\textbf{Construct validity}} concerns whether the theoretical constructs are correctly interpreted and measured. All researchers agreed on the questions and scope of the study. The principal researcher recorded and transcribed the interviews. He listened and only clarified questions upon the interviewees' request to prevent the researcher's interpretation from impacting their responses.

\textit{\textbf{Internal validity}} refers to whether the results adhere to the data. To analyze the data, we conducted a qualitative thematic analysis to categorize the results inductively and report them in the study. 

\textit{\textbf{External validity}} concerns whether the results' generality claims are justifiable. The current study's limited sample size is statistically irrelevant. The study aimed to collect expert views regarding sustainability in agile software development within the Finnish software industry, so the statistical approach was not the aim.

\textit{\textbf{Reliability}} concerns whether the study is replicable if other researchers conduct it. We followed a structured interview process to mitigate this threat, meaning the questions were the same in all the interviews.

%% file: sections/roadmap.tex
\section{Software Engineering Roadmap Contributions}
The adoption of sustainability concepts into software engineering has been a topic of increasing interest in recent years, especially in the Software development life cycle context \cite{mourao_green_2018}. Agile methods have gained traction over the past several years, and their use, interest, and discussions have expanded rapidly, as have the anecdotal indications of their efficacy in particular environments and within specific project types \cite{lindvall2004agile}. Due to their popularity, implementing sustainability within these methodologies can have a broader impact on sustainability within the software industry.

Sustainability has piqued significant interest in the software industry, particularly in the context of agile methods, as evidenced by the views expressed by experts in this study. This indicates that the software industry's next step is integrating sustainability values into its agile software development processes. We propose three milestones that, if implemented, could significantly enhance the sustainability of agile software development, thereby contributing to the software engineering roadmap 2030:
\begin{enumerate}
    \item \textbf{Educating in sustainability:} Software practitioners need education regarding sustainability to increase sustainability knowledge and awareness. Below, we listed two areas where we propose to include sustainability topics.
    \begin{enumerate}
        \item \textbf{Staff training:} Including sustainability within staff training programs to increase employees' knowledge of sustainability is a step in the right direction.
        \item \textbf{Educational institutions:} Future software experts start at educational institutions such as universities, and including sustainability in these institutions within the software and agile topics can positively impact sustainability knowledge and awareness.
    \end{enumerate}
    \item \textbf{Sustainability tools and frameworks:} Tools and frameworks must be developed to address sustainability challenges and impacts within agile software development.
      \begin{enumerate}
       \item \textbf{Frameworks:} It is essential to develop frameworks that aid practitioners in integrating sustainability into agile software development. The frameworks should consider five aspects of sustainability, including environmental, economic, individual, social, and technical, and include specific and detailed actions and principles that help practitioners incorporate sustainability in agile software development.
        \item \textbf{Measurement tools:} Tools are needed to measure, for example, the energy efficiency of software development libraries, frameworks, and tools to aid developers in achieving sustainability in agile software development. 
       
    \end{enumerate}
    \item \textbf{Energy Efficiency Measurement:} Libraries' energy efficiency in software development must be measured, and the results must be shared to aid experts in choosing the best option to develop sustainable software in agile software development teams. Policies are needed to urge organizations to measure the energy efficiency of the libraries they develop and share their information. 

\end{enumerate}

%% file: sections/conclusion.tex
\section{Conclusion}

The current study utilized an interview approach to explore sustainability in agile software development within the Finnish software
industry. The study reflects the growing interest in sustainability within the industry and emphasizes the need to raise awareness about it. The study also brings to light the importance of various sustainability dimensions in agile software development and how they are interconnected. Moreover, it suggests that incorporating sustainable practices into Scrum software development could have a broader impact on the Finnish software industry. 

This study underscores the practical implications of integrating sustainability into software education and developing practices to tackle sustainability challenges in the software industry. It provides valuable insights for researchers to collaborate with the software industry in developing frameworks and tools to address sustainability in agile software development. The study outlines three key milestones: educating in sustainability, developing sustainability tools and frameworks, and measuring the energy efficiency of software libraries to contribute to the software engineering roadmap 2030.

%% file: sections/appendix.tex
\section{Appendix}
\begin{enumerate}
    \item What is your job role in your organization?
    \item How many years of experience do you have in your organization and in software development in general? 
    \item What industry best describes your organization?
    \item What is the size of the organization you work in?
    \item What is your perception of sustainability in software development?
    \item How is sustainability considered in your organization on a scale of 1 to 5? (1 least consideration, five highest consideration)
    \item What agile framework or methods are currently being used in your company?
    \item Which sustainability objectives may your organization have that are relevant to agile practice? 
    \item What are the top three sustainability dimensions that are most relevant to agile software? 
    \begin{enumerate}
    \item Economic
    \item Environmental
    \item Individual
    \item Social
    \item Technical
    \end{enumerate}
    \item What are the challenges in agile software development affecting sustainability in your organization?
    \item How do the challenges impact the agile software development process in your organization?
    \item In your opinion, what can be done to improve sustainability in agile software development?
\end{enumerate}